\documentclass[conference]{IEEEtran}
\IEEEoverridecommandlockouts

\makeatletter
\def\ps@IEEEtitlepagestyle{
  \def\@oddfoot{\mycopyrightnotice}
  \def\@evenfoot{}
}
\makeatother
\def\mycopyrightnotice{
  \parbox{\textwidth}{
    {\footnotesize \url{https://doi.org/10.1109/QCE57702.2023.10206}~\copyright~2023 IEEE.  
  Personal use of this material is permitted.
  Permission from IEEE must be obtained for all other uses, in any current or future media, 
  including reprinting/republishing this material for advertising or promotional purposes, 
  creating new collective works, for resale or redistribution to servers or lists, 
  or reuse of any copyrighted component of this work in other works.}
  }
  \gdef\mycopyrightnotice{}
}

\usepackage[colorlinks=false,hidelinks]{hyperref}
\usepackage{varioref}
\usepackage{amsmath,amssymb,amsfonts}
\usepackage{algorithmic}
\usepackage{graphicx}
\usepackage{textcomp}
\usepackage{xcolor}
\usepackage{dblfloatfix}
\usepackage{cleveref}
\usepackage{subcaption}
\usepackage{orcidlink}
\usepackage[braket, qm]{qcircuit}

\begin{document}

\title{Comparing Quantum Service Offerings\\
\thanks{ This work was funded in part by the BMWK projects \textit{SeQuenC} (01MQ22009B), \textit{EniQmA} (01MQ22007B), and \textit{PlanQK} (01MK20005N). }
}
\author{
\IEEEauthorblockN{Julian Obst\orcidlink{0000-0002-1898-2167}, Johanna Barzen\orcidlink{0000-0001-8397-7973}, Martin Beisel\orcidlink{0000-0003-2617-751X}, Frank Leymann\orcidlink{0000-0002-9123-259X}, Marie Salm\orcidlink{0000-0002-2180-250X}, and Felix Truger\orcidlink{0000-0001-6587-6431}
    \IEEEauthorblockA{University of Stuttgart, Institute of Architecture of Application Systems, Universitätsstraße 38, Stuttgart, Germany
    \\\{obst, barzen, beisel, leymann, salm, truger\}@iaas.uni-stuttgart.de
		}
	}
}

\maketitle

\begin{abstract}
With the emergence of quantum computing, a growing number of quantum devices is accessible via cloud offerings.
However, due to the rapid development of the field, these quantum-specific service offerings vary significantly in capabilities and requirements they impose on software developers.
This is particularly challenging for practitioners from outside the quantum computing domain who are interested in using these offerings as parts of their applications.
In this paper, we compare several devices based on different hardware technologies and provided through different offerings, by conducting the same experiment on each of them.
By documenting the lessons learned from our experiments, we aim to support developers in the usage of quantum-specific offerings and illustrate the differences between predominant quantum hardware technologies. 
\end{abstract}


\section{Introduction}
\label{sec:intro}
Quantum computing promises a computational advantage over its classical counterpart in various fields, such as optimization and chemistry~\cite{Harrigan2021optimization,cao2018qcpotential}.
However, the capabilities of current \textit{Noisy Intermediate-Scale Quantum~(NISQ)} devices are still limited by low numbers of qubits and high error rates~\cite{Leymann2020_QuantumAlgorithmsBitterTurth,Preskill2018}.
To improve their quality, various approaches to implement the qubits of a quantum device, e.g., trapped ions~\cite{Bruzewicz2019TrappedIon} or superconducting electrical circuits~\cite{Preskill2018}, are currently being explored.
To enable easy access to quantum devices, they are typically provided via the cloud by so-called \emph{Quantum Computing as a Service~(QCaaS)} offerings~\cite{Leymann2020_QuantumCloud}.

However, the heterogeneity of these QCaaS offerings, as well as differences of the available quantum devices, impose numerous problems to quantum software developers.
These include designing the quantum algorithm to meet hardware capabilities, the implementation and execution of the resulting quantum circuits, and the evaluation of execution results.
Overcoming these issues is particularly difficult for practitioners from outside the quantum domain, as existing offerings and tools are tailored for domain experts.
To prepare unfamiliar developers for the problems imposed by current quantum devices and QCaaS offerings in practice,
we execute the well-known \emph{Quantum Approximate Optimization Algorithm~(QAOA)}~\cite{fahri2014qaoa} on four devices provided through different QCaaS offerings and (i)~describe the issues we encountered during this process.
Moreover, we (ii)~evaluate and compare the execution results by employing the mean absolute difference as a metric to assess their noisiness.

\section{Background and Related Work} 
\label{sec:background}

There exist different physical realizations of qubits such as ion traps, quantum dots, or superconducting qubits.
Our experiments use quantum devices with superconducting qubits or trapped ions, being common quantum technologies~\cite{whitfield2022quantum}.
\looseness=-1

We evaluate QAOA~\cite{fahri2014qaoa} and consider an instance of the \emph{MaxCut} problem, where the goal is to divide a weighted graph's vertices into two partitions to maximize the total weight of edges with endpoints in different partitions.
MaxCut was considered in the seminal QAOA paper~\cite{fahri2014qaoa} and is widely used as a benchmark since~\cite{baker2022wasserstein,Pelofske2021fairsampling}.
QAOA is a \emph{Variational Quantum Algorithm (VQA)}, i.e., it employs a classical optimizer to train a parameterized quantum circuit~\cite{Cerezo2021vqa}.
The circuit comprises two sub-circuits parameterized by angles $\gamma,\beta$, executed subsequently.
The optimizer iteratively improves the values of $\gamma,\beta$, based on previous results.
QAOA's depth can be adjusted via a hyperparameter $p$, specifying the number of repetitions of the sub-circuit pair. 
Limitations of NISQ devices require low $p$, e.g., $p=1$ or $p=2$ (depth-1, depth-2 QAOA), yet, QAOA can provide meaningful results, making it suitable for the NISQ era.
The number of adjustable parameters increases with $p$, e.g., depth-2 QAOA, optimizes $\gamma_1,\gamma_2,\beta_1,\beta_2$.
With increasing $p$, the algorithm converges towards an optimum~\cite{fahri2014qaoa}.
\looseness=-1

Pelofske~et~al.~\cite{Pelofske2021fairsampling} and Baker and Radha~\cite{baker2022wasserstein} benchmark quantum devices using QAOA.
They compare the balance of the optimal solution and perform mean-variance portfolio optimization, respectively.
These benchmarks focus on the accuracy of results, our work compares devices through experiments and focuses on the development, access, and execution processes using QCaaS offerings, outlining encountered issues and observations to support developers with knowledge about typical pitfalls and obstacles.
Unlike Vietz et al.~\cite{Vietz2021_QuantumSoftwareEngineeringChallenges}, who focus on engineering quantum algorithms for the cloud, i.e., composition and integration, our work concentrates on the obstacles during implementation for different QCaaS offerings and comparing experimental results from different devices.
\looseness=-1

\begin{figure}[h]
   \centering
   \includegraphics[trim=0 0 6.5cm 0, clip, width=0.5\textwidth]{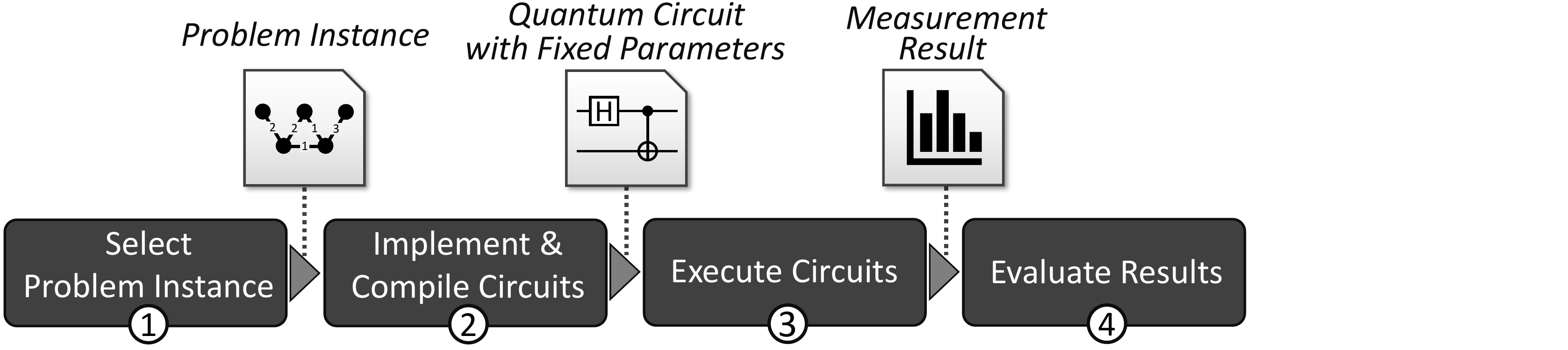}
   \caption{Procedure of developing, executing, and evaluating QAOA circuits.}
   \label{fig:experimentDesign}
\end{figure}

\section{Experiments and Encountered Issues}
\label{sec:experiments}

In our experiments, we follow the process depicted in~\Cref{fig:experimentDesign} to examine QAOA for MaxCut on a selection of quantum devices and evaluate the optimization landscape of QAOA's variational parameters.
This approach enables a more objective comparison of devices, avoiding uncertainties from parameter optimization.
The optimization landscape is a visualization of the objective function.
Here, the objective function is the expected value of a cut that is the result of a measurement.
While these landscapes seem similar to plots of complex analysis \cite{wegert2011phaseplots}, the landscapes here are not complex functions.
To create the landscape, we sample equidistant points in the parameter space of depth-1 and depth-2 QAOA.
For each sampled point (1000 shots), we evaluate the expectation value in order to interpolate the optimization landscape.
Due to resource limitations, we restrict the sampled parameter space to $(\beta,\gamma)\in([0, \frac{\pi}{2}], [0,\pi])$, where each sample is $\frac{\pi}{20}$ apart.
In the depth-2 case, we fix $\beta_1, \gamma_1$ to their optimal values found in the depth-1 case to visualize the 2-dimensional $(\beta_2, \gamma_2)$-optimization landscape.
We thereby follow the idea of a layer-wise parameter initialization that can be seen as a warm-start of depth-2 QAOA utilizing the result from \mbox{depth-1} QAOA~\cite{Truger2023_WarmStarting}.
The experiments were run on four devices, all manufactured by different vendors, and a simulator for comparison with error-free results.
Due to confidentiality reasons and the limited scope of our experiments, the names of the offerings are omitted to avoid premature conclusions.

\subsection{Selecting the Problem Instance}

In the first step, shown in~\Cref{fig:experimentDesign}, an instance of the MaxCut problem, i.e., a graph, needs to be selected.
As the complexity of typical real-world problems exceeds the capabilities of current quantum devices, we chose a small graph (see \Cref{fig:experimentDesign}), which is suitable for NISQ devices.

\begin{figure*}[t]
    \centering
    \Qcircuit @C=0.47em @R=0.4em @!R { \\
	 	\nghost{{q}_{0} :  } & \lstick{{q}_{0} :  } & \gate{H} \barrier[0em]{4} & \qw & \targ & \gate{R_Z\,(-3\gamma)} & \targ & \qw & \qw & \qw & \qw & \qw & \qw & \qw & \qw & \qw & \qw & \qw & \qw \barrier[0em]{4} & \qw & \gate{R_X\,(2\beta)} & \qw & \qw\\
	 	\nghost{{q}_{1} :  } & \lstick{{q}_{1} :  } & \gate{H} & \qw & \ctrl{-1} & \qw & \ctrl{-1} & \targ & \gate{R_Z\,(-1\gamma)} & \targ & \targ & \gate{R_Z\,(-1\gamma)} & \targ & \qw & \qw & \qw & \qw & \qw & \qw & \qw & \gate{R_X\,(2\beta)} & \qw & \qw\\
	 	\nghost{{q}_{2} :  } & \lstick{{q}_{2} :  } & \gate{H} & \qw & \qw & \qw & \qw & \ctrl{-1} & \qw & \ctrl{-1} & \qw & \qw & \qw & \targ & \gate{R_Z\,(-2\gamma)} & \targ & \qw & \qw & \qw & \qw & \gate{R_X\,(2\beta)} & \qw & \qw\\
	 	\nghost{{q}_{3} :  } & \lstick{{q}_{3} :  } & \gate{H} & \qw & \qw & \qw & \qw & \qw & \qw & \qw & \qw & \qw & \qw & \qw & \qw & \qw & \targ & \gate{R_Z\,(-2\gamma)} & \targ & \qw & \gate{R_X\,(2\beta)} & \qw & \qw\\
	 	\nghost{{q}_{4} :  } & \lstick{{q}_{4} :  } & \gate{H} & \qw & \qw & \qw & \qw & \qw & \qw & \qw & \ctrl{-3} & \qw & \ctrl{-3} & \ctrl{-2} & \qw & \ctrl{-2} & \ctrl{-1} & \qw & \ctrl{-1} & \qw & \gate{R_X\,(2\beta)} & \qw & \qw\\
    }
    \caption{QAOA circuit of our instance.
    The final measurement operations are omitted.
    Note that each edge of the graph is represented as an $R_Z$-gate between two CNOT-gates.}
    \label{fig:circuit}
\end{figure*}
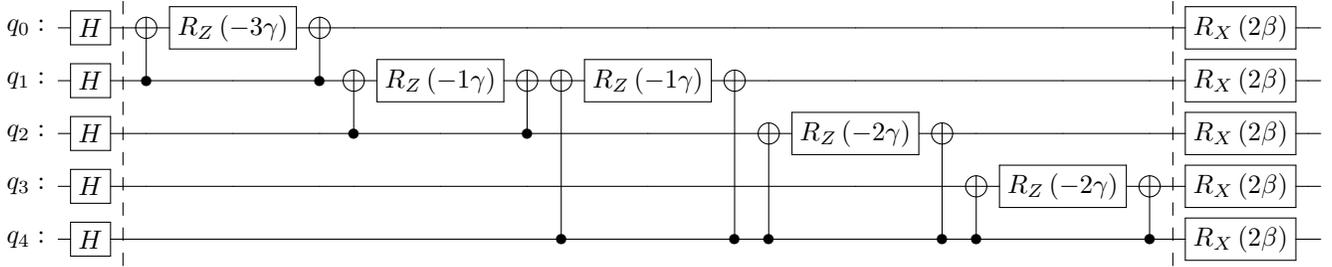

\smallskip
\noindent
\textbf{Connectivity Requirements.}
The structure of the problem instance already imposes issues, as it requires certain qubit connections for the execution on a device.
For the MaxCut problem tackled with QAOA, the targeted graph directly shows the required connectivity, since each node of the graph will be mapped to one qubit and each edge of the graph introduces two-qubit gates between the adjacent qubits.
The resulting circuit is depicted in~\Cref{fig:circuit}.
Our problem instance requires a moderate connectivity with 5 out of 10 possible connections between 5 nodes, and is thus feasible for devices ranging between low and full connectivity.

\subsection{Implement \& Compile Circuits}
\label{subsec:implement_transpile_circuits}
In the second step of~\Cref{fig:experimentDesign}, the QAOA circuits are implemented and compiled for the quantum device they shall be executed on. 
In the following, we discuss the issues encountered during the implementation and compilation. 

\smallskip
\noindent
\textbf{Heterogeneous Ecosystems and Formats.}
The lack of an established vendor-independent language for implementing quantum algorithms complicates their development for developers who do not want to restrict themselves to a specific ecosystem, risking a vendor lock-in.
Although there are tools to translate between different formats~\cite{qconvert2023,sivarajah2021tket}, they often only support specific releases of the frequently updated quantum SDKs and do not always provide a semantically equivalent result~\cite{kharkov2022arline}.
Thus, despite the availability of various translation tools, we had to manually adapt our implementation for individual QCaaS offerings.

\smallskip
\noindent
\textbf{Compilation Quality.}
\label{subsec:transpilation_quality}
As most quantum devices have a restricted connectivity between their qubits and support only a subset of quantum gates natively, quantum circuits must be compiled for the target device~\cite{Leymann2020_QuantumAlgorithmsBitterTurth}.
This process is NP-hard~\cite{siraichi2018compilationNPHard}; thus, the compilation results can differ significantly, e.g., in circuit depth or the number of two-qubit gates~\cite{Salm2021_CompilerComparison}.
Most offerings give insight into the compiled circuits, such that the compilation process can be repeated multiple times, enabling their comparison and selection for execution.
However, one of the offerings used for our experiments encrypts the compilation results, making an analysis and comparison infeasible.
\looseness=-1

\subsection{Execute Circuits}
\label{subsec:execution}

After implementing the quantum circuit addressing the chosen MaxCut problem and compiling it for the different devices, it can be executed. 
However, the execution step imposes various offering-specific issues.

\smallskip
\noindent
\textbf{Execution Endpoints.}
The execution endpoints of the different QCaaS offerings differ in many aspects:
First, some offerings do not support batching multiple circuits, e.g., for sampling a parameter space, as described above, into a single execution job.
Hence, developers must process all circuit execution requests and responses individually, making it an error-prone process.
Another issue is that one offering uses different formats for simulated and real quantum devices.
Hence, developers cannot be certain that their implementations tested with a simulator will also work on a costly quantum device.
Another inconvenience that we observed when executing circuits using pay-per-use QCaaS offerings, is the difficulty of estimating the execution costs, particularly for VQAs, as the number of iterations varies.

\smallskip
\noindent
\textbf{Execution Time.}
Waiting times significantly exceeded the execution times in our experiments.
Thus, different options for accessing the devices must be considered to reduce waiting times.
For example, queuing can lead to waiting times of multiple hours.
However, some offerings also provide exclusive time slot reservations or high-priority access.
In the context of reservations, we noticed that it is important to efficiently use the usually short time slots of a few minutes, e.g., by pre-compiling circuits and avoiding extensive classical processing between circuit executions.
One device that is still in experimental stage and human-operated could not be automatically accessed via an execution endpoint.
\looseness=-1

\subsection{Evaluate Results}
After executing the circuit and obtaining the measurement results, the final step shown in~\Cref{fig:experimentDesign} entails their evaluation.

\smallskip
\noindent
\textbf{Result Format.}
The offerings' responses differ significantly.
In particular, the measurement results are returned in different formats, e.g., an aggregated list of measured bit strings and their frequency, or the bit strings measured for each shot where the user has to aggregate the data by themselves.
Additionally, offerings annotate the results with different metadata, e.g., the name and version of the quantum device.
The amount and detail of metadata differs across the offerings, hindering a comparison among them.
For example, due to missing information about the execution times of some offerings, we were not able to directly compare them.
Another pitfall is that some offerings return bit strings in reversed order, requiring careful interpretation of the results by developers.

\section{Experiment Results}
\begin{figure*}
    \centering
    \includegraphics[trim=0 0 0 0,clip,width=0.85\textwidth]{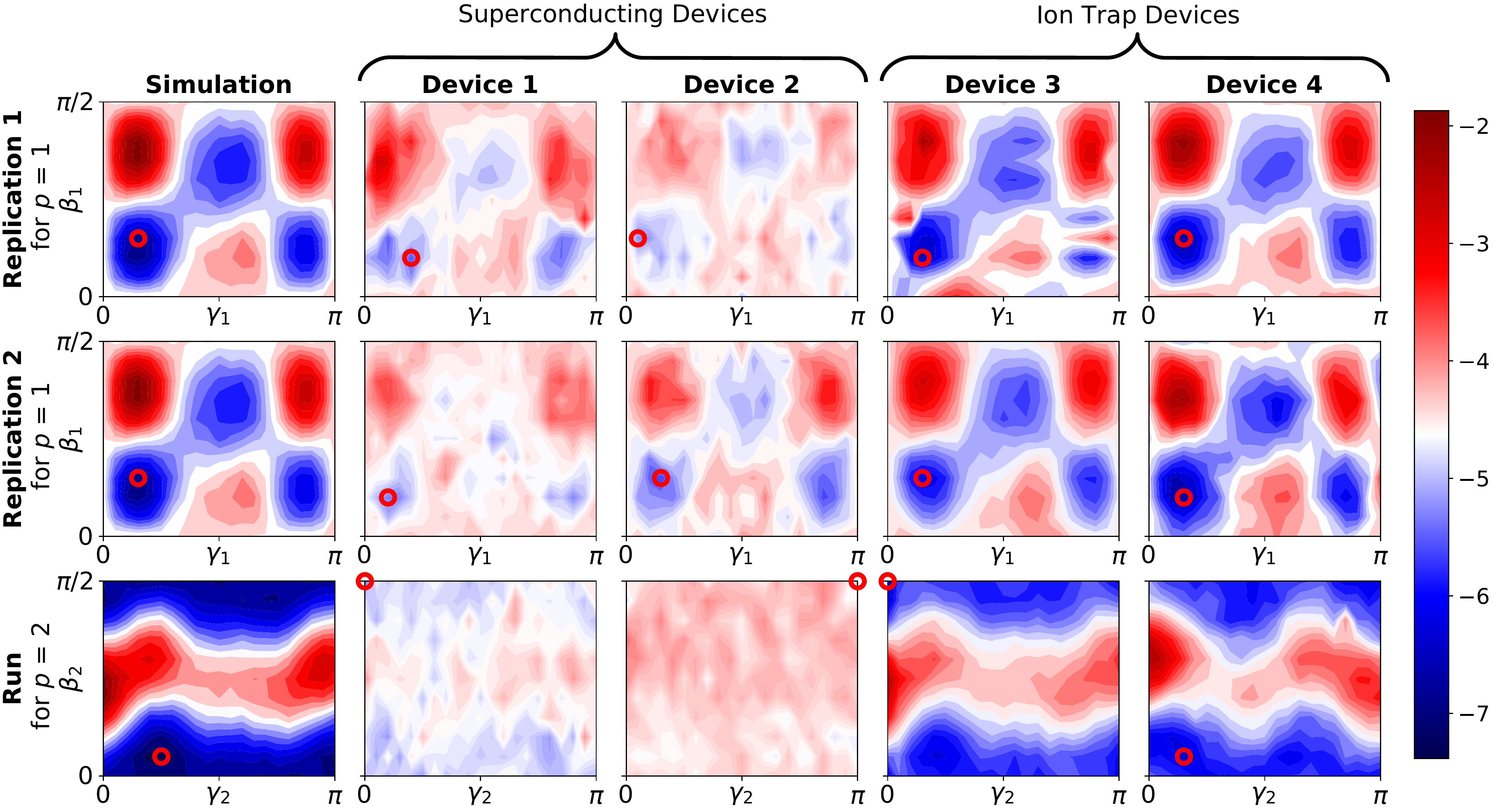}
    \caption{Plotted energy expectation in the ($\gamma_1, \beta_1$) and ($\gamma_2$, $\beta_2$) parameter spaces.
    Measured minima are marked by red circles.
    The first two rows show results for depth-1 QAOA, replicated on different days.
    The bottom row shows results for depth-2 QAOA, with parameters $\beta_1,\gamma_1$ fixed to optimal values from replication 2.
    }
    \label{fig:gridsearch}
    \vspace{-0.7em}
\end{figure*}

The results of our experiments are illustrated in~\Cref{fig:gridsearch}.
Their quality varies significantly across the four devices, and particularly across devices based on different hardware technologies.
The contour plots depict the objective function for QAOA in the so-called optimization landscape.
As explained above, it is a visualization of the function that the optimizer tries to minimize.
Columns correspond to simulation and the different devices, and rows to two replications of depth-1 and one run for depth-2 QAOA.
Low (blue) values are desirable, as they resemble lower energy, corresponding to better solutions of the MaxCut problem.
In contrast, high (red) values indicate unfavorable parameter values.

\begin{figure}
    \centering
    \includegraphics[trim=0 10 20 0, clip, width=.5\textwidth]{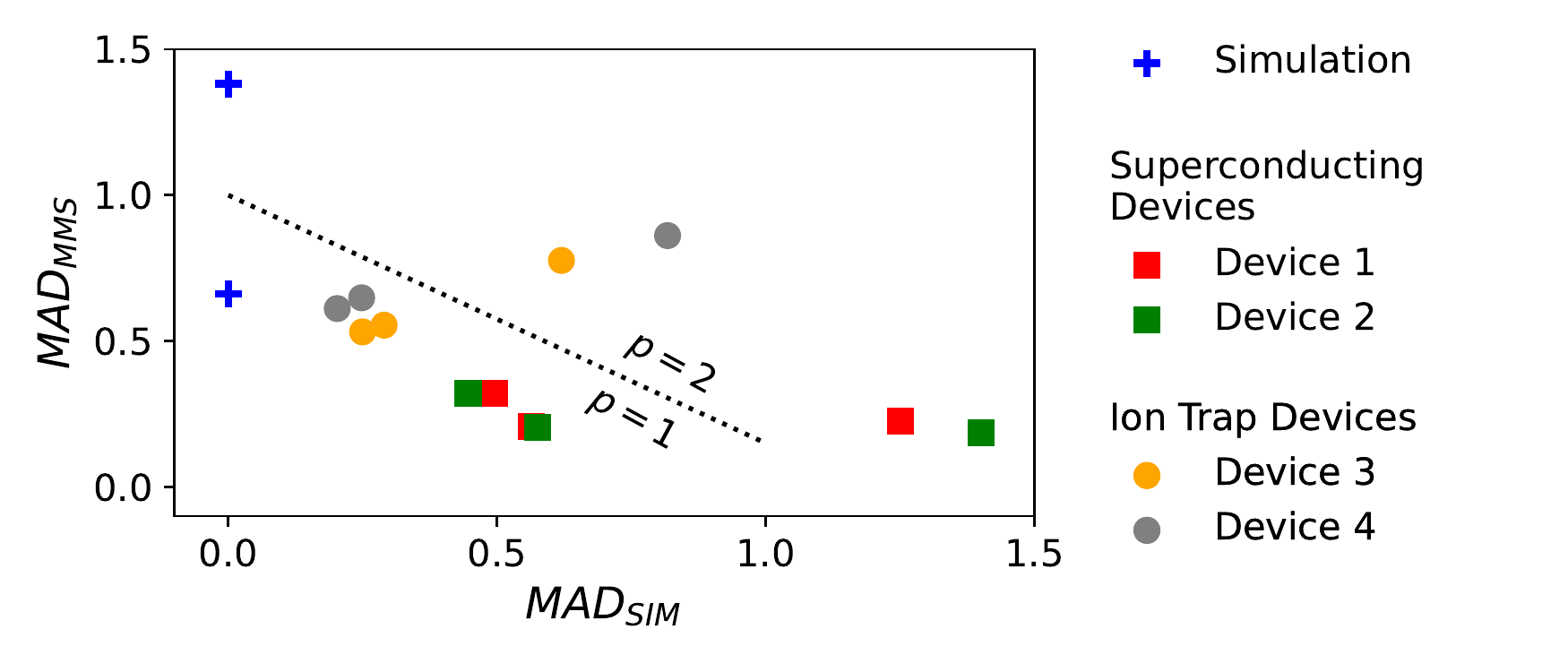}
    \caption{MADs of the energy expectation values plotted in~\Cref{fig:gridsearch}.
    A low MAD\textsubscript{SIM} and a high MAD\textsubscript{MMS} are favorable.}
    \label{fig:MAD}
\end{figure}

The fact that result quality can change over time is demonstrated by the two replications for depth-1 QAOA per offering.
Thus, repeating experiments at a later point with updated device calibrations can affirm the validity of results.
Evidently, various error sources, deteriorate the results significantly compared to the simulation in the leftmost column of~\Cref{fig:gridsearch}.
These results also illustrate the impact of the circuit depth, as the landscapes for depth-2 QAOA in the bottom row appear less similar to the simulation compared to those for depth-1 in the upper rows.
These insights are underlined by the \emph{Mean Absolute Differences~(MADs)} shown in~\Cref{fig:MAD}, which are calculated by averaging the absolute differences between the samples of two landscapes:
\begin{equation*}
    \mathrm{MAD}(E_1,E_2) = \frac{1}{|\Gamma| \cdot |B|}\sum_{(\gamma, \beta) \in \Gamma \times B} \left|E_1(\gamma, \beta) - E_2(\gamma, \beta)\right|
\end{equation*}
$\Gamma=\{0, \frac{\pi}{20}, \dots, \frac{19\pi}{20}, \pi\}$ and $B=\{0, \frac{\pi}{20}, \dots, \frac{9\pi}{20}, \frac{\pi}{2}\}$ are the samples for $\gamma$ and $\beta$ respectively, and $E_1,E_2$ are the expectation values of the objective function, also called energy, for two different landscapes.
The MAD is an intuitive measure for the difference of two ordered sets.
It can be used to assess the performance of a model~\cite{Willmott2005}.
The MAD\textsubscript{SIM} serves as a similarity measure between landscapes obtained from the quantum devices and the simulation, whereas the MAD\textsubscript{MMS} is a similarity measure between landscapes obtained from the devices and the \emph{Maximally Mixed State (MMS)}, i.e., the mixture where every basis state is equally probable.
The MMS resembles a random sampling of bit strings, therefore, the MAD\textsubscript{MMS} is an indicator of noise~\cite{Wang2021Noise}. 

Both~\Cref{fig:gridsearch,fig:MAD} show a clear difference between the two types of devices.
While all devices found a minimum in relative proximity to the true minimum in the depth-1 case, the landscapes produced by the trapped ion devices resemble the simulation much closer.
This becomes even more evident in the depth-2 case, where the result of device 2 is dominated by noise.
As the MAD\textsubscript{SIM} in~\Cref{fig:MAD} shows, the increased QAOA depth also led to a reduced similarity to the simulation.
For the ion trap devices, the MAD\textsubscript{MMS} in~\Cref{fig:MAD} increased slightly with the QAOA depth.
A greater MAD\textsubscript{MMS} indicates a less noisy result.
Evidently, ion trap devices performed better in our experiments.
A possible explanation is the graph density of the problem instance.
Pelofske et al.~\cite{Pelofske2021fairsampling} also found that trapped ion devices performed better for instances requiring a high connectivity.
However, our experiments only cover one use case, hence, other circuit types may lead to different results.
\looseness=-1

\section{Summary and Future Work}
\label{sec:summary}
In this work, we documented the lessons learned using different QCaaS offerings.
To this end, we implemented and executed the same experiment, addressing the MaxCut problem with QAOA, for four quantum devices based on ion-trap and superconducting qubits. 
The heterogeneity of the QCaaS offerings complicated the development and execution.
In particular, differences in formats, access options, and limitations, e.g., regarding metadata, had to be considered.
Additionally, noise and errors of current quantum devices deteriorated the results.
However, the impact differs depending on the underlying hardware technology.
In our experiments, the results obtained from superconducting devices were significantly more erroneous compared to those of ion trap devices.
\looseness=-1

In future work, we plan to address the heterogeneity of QCaaS offerings by means of a unified access layer, facilitating device-agnostic development and execution.
Furthermore, we will expand our experiments to more use cases, quantum algorithms, quantum devices, and hardware technologies.
\looseness=-1

\bibliographystyle{IEEEtran}
\bibliography{bibliography}

\end{document}